\documentclass[twocolumn,trackchanges,twocolappendix]{aastex631}
\usepackage{natbib}
\usepackage{epsfig}
\usepackage{graphicx}
\usepackage{subfigure}
\usepackage{float}
\usepackage{amsmath}
\usepackage{color}
\usepackage{amssymb}
\usepackage[utf8]{inputenc}
\usepackage{apjfonts}
\usepackage{hyperref}

\newcommand{\PMO}{Purple Mountain Observatory, Chinese Academy of Sciences, Nanjing 210023, China}
\newcommand{\GZNU}{School of Physics and Electronic Science, Guizhou Normal University, Guiyang 550001, China}

\begin{document}

\title{Detecting Population III Gamma-Ray Bursts with Einstein Probe and Space-Based Multi-band Astronomical Variable Objects Monitor}

\correspondingauthor{Jun-Jie Wei, Qing-Bo Ma, and Xue-Feng Wu}
\email{jjwei@pmo.ac.cn, maqb@gznu.edu.cn, xfwu@pmo.ac.cn}

\author[0000-0003-0162-2488]{Jun-Jie Wei}
\affiliation{\PMO}

\author[0000-0001-9493-4565]{Qing-Bo Ma}
\affiliation{\GZNU}

\author[0000-0002-6299-1263]{Xue-Feng Wu}
\affiliation{\PMO}

\begin{abstract}
High-redshift gamma-ray bursts (GRBs), putative counterparts of massive, low-metallicity
Population III (Pop III) stars, are a promising probe of the first stars. We assess the
detectability of these Pop III GRBs using a metallicity-based progenitor criterion and
cosmological $N$-body/hydrodynamical simulations with three distinct Pop III initial
mass functions (IMFs), focusing on the capabilities of the Wide-field X-ray Telescope
(WXT) aboard the Einstein Probe (\emph{EP}) and the coded-mask gamma-ray imager (ECLAIRs)
aboard the Space-based multi-band astronomical Variable Objects Monitor (\emph{SVOM}).
Our population synthesis model, calibrated to \emph{Swift} data, predicts the following
Population II/I (Pop II/I) GRB detection rates at $z>6$: $\sim2.4\,\mathrm{events\,yr^{-1}}$
for \emph{EP}/WXT and $\sim0.9\,\mathrm{events\,yr^{-1}}$ for \emph{SVOM}/ECLAIRs.
For the IMF with very massive first stars ($\mathrm{100\textrm{--}500\,M_\odot}$),
we derive upper limits on the Pop III GRB rate at $z>6$ of $<0.06\,\mathrm{events\,yr^{-1}}$
(\emph{EP}/WXT) and $<0.13\,\mathrm{events\,yr^{-1}}$ (\emph{SVOM}/ECLAIRs), based on
the absence of confirmed Pop III progenitors in \emph{Swift} bursts at $z>5.5$. Our results
indicate that while Pop III GRBs are subdominant to Pop II/I GRBs at $z<10$, their fractional
contribution rises significantly with redshift, reaching $\sim8\%$ ($\sim34\%$) at $z>10$
and $\sim28\%$ ($\sim68\%$) at $z>16$ for \emph{EP}/WXT (\emph{SVOM}/ECLAIRs). This trend
is systematically enhanced in the other two IMF models, which adopt a lower stellar mass
range of $\mathrm{[0.1,\,100]\,M_\odot}$. We conclude that detecting Pop III GRBs at high redshifts
is a realistic prospect, and any GRB detected at $z>16$ is most likely of Pop III origin.
\end{abstract}


\section{Introduction}
\label{sec:intro}
Population III (Pop III) stars, the first stellar generation, formed from metal-free primordial gas (hydrogen
and helium) and brought an end to the cosmic dark ages (see \citealt{2004ARA&A..42...79B,2011ARA&A..49..373B,2023ARA&A..61...65K}
for reviews). These stars played a pivotal role in early cosmic evolution: their intense ultraviolet (UV)
radiation reionized the Universe (e.g., \citealt{1997ApJ...486..581G,2000ApJ...528L..65T}), and their
supernovae (SNe) enriched the intergalactic medium with heavy elements, thereby influencing subsequent
stellar generations (e.g., \citealt{2002ApJ...567..532H,2004ApJ...605..579Y}). Cosmological simulations
within the $\Lambda$CDM model suggest Pop III stars formed at redshift $z\gtrsim30$, dominating star formation
until $z\sim15-20$ prior to the emergence of metal-enriched Pop II stars (e.g., \citealt{2022ApJ...936...45H}).
Despite their importance, quantifying the formation history of Pop III stars remains highly uncertain
\citep{2001PhR...349..125B}. Direct observational probes are not yet feasible \citep{2020ApJ...904..145S},
and current numerical models (e.g., \citealt{1999ApJ...527L...5B,2002ApJ...564...23B,2002Sci...295...93A})
are limited by resolution and physical complexity. Indirect methods, such as studying the chemical imprints
of the first stars in ancient, metal-poor stars \citep{2006ApJ...641....1T}, depend on future large-scale
spectroscopic surveys \citep{2021MNRAS.502....1J} for validation. Given these limitations, determining
the initial mass function (IMF) and star formation rate (SFR) of the first stars constitutes a fundamental
challenge in modern astrophysics \citep{2025arXiv250803689M}.

Most long-duration gamma-ray bursts (GRBs), defined by durations exceeding two seconds,
are linked to the collapse of massive stars and the subsequent accretion onto newborn
black holes (BHs; e.g., \citealt{1993ApJ...405..273W,1998ApJ...494L..45P,2006ARA&A..44..507W}).
Due to their extreme brightness, GRBs are detectable at very high redshifts, as evidenced by
GRB 090423 at $z=8.2$ \citep{2009Natur.461.1258S,2009Natur.461.1254T} and GRB 090429B at $z\sim9.4$
\citep{2011ApJ...736....7C}. Moreover, theoretical models suggest that some high-$z$ GRBs could
originate from the collapse of massive Pop III stars \citep{2010ApJ...715..967M,2011ApJ...726..107S,2011ApJ...731..127T}.
Therefore, high-$z$ GRBs provide a powerful probe for directly studying and constraining
the properties and formation history of the first stars \citep{2000ApJ...536....1L,2002ApJ...575..111B,
2011MNRAS.416.2760C,2011MNRAS.414..847S,2012ApJ...760...27W,2015NewAR..67....1W,2015JHEAp...7...35S,2016SSRv..202..159T}.
Indirectly, the metal enrichment from Pop III stars can be gauged from absorption features
in later Pop II GRB spectra, which reveal an environment enriched by the first SNe \citep{2012ApJ...760...27W}.
A consensus across diverse models indicates that Pop III GRBs would be exceptionally energetic,
with isotropic-equivalent energies orders of magnitude greater than those of Pop II events.
For instance, \cite{2011ApJ...731..127T} estimated energies of $\sim10^{56}-10^{57}\,\mathrm{erg}$
for such bursts, which would make them detectable even at the highest redshifts. A prolonged
prompt emission lasting up to $\sim 10^4\,\mathrm{s}$ is also predicted in most models.
These traits, however, are observationally indistinguishable from those of lower-redshift
ultra-long GRBs \citep{2014ApJ...781...13L}, which likely stem from Pop II blue supergiants
\citep{2013ApJ...778...67N,2018ApJ...859...48P}, and thus cannot be considered unique signatures
of Pop III progenitors. Definitive confirmation may instead require detecting the absence of
metal absorption lines in afterglow spectra, indicating a metallicity below the critical value.
Yet, achieving the requisite spectroscopic sensitivity for such measurements remains challenging
even for 30-m class telescopes. Given the lack of unambiguous spectroscopic identifiers,
radio afterglows provide a key diagnostic for Pop III GRBs. Their unique energetics are predicted
to generate exceptionally powerful radio emission \citep{2000ApJ...540..687C,2011ApJ...731..127T,2013MNRAS.435.2543G}.
As demonstrated by \cite{2013MNRAS.435.2543G}, Pop III radio afterglows are distinguishable from
Pop II events by reaching much higher peak fluxes at later times, occupying a unique area
in the peak time–flux plane.

Despite promising identification methods, Pop III GRBs have not yet been unambiguously observed.
A consensus holds that GRBs detected at $z>6$ represent the high-redshift tail of the Pop II/I
distribution, not Pop III events \citep{2014ApJ...781....1L,2022A&A...665A.125R,2025A&A...695A.239B}.
This non-detection provides a key constraint on the theoretical
event rate of Pop III GRBs \citep{2006ApJ...642..382B,2011MNRAS.416.2760C,2011ApJ...731..127T,
2011A&A...533A..32D,2014MNRAS.439.3520M,2014ApJ...787...91M,2015MNRAS.449.3006M,2025arXiv250803689M},
prompting numerical simulations that quantify the expected detection rate for missions like \emph{Swift}
(e.g., \citealt{2011MNRAS.416.2760C,2015MNRAS.449.3006M}). A major uncertainty in these predictions
stems from the unknown Pop III IMF. Many simulations suggest a top-heavy Pop III IMF with characteristic
stellar masses of hundreds of solar masses \citep{1998MNRAS.301..569L,2002Sci...295...93A,2004ApJ...605..579Y},
while a standard low-mass IMF (with masses well below $\sim\mathrm{100\,M_\odot}$) is also plausible \citep{2007ApJ...667L.117Y,2008A&A...490..769C,2010MNRAS.405..177S}. These competing hypotheses introduce
variations of over an order of magnitude in the estimated Pop III SFR
\citep{2010MNRAS.407.1003M,2011MNRAS.414.1145M,2011MNRAS.415.3021M}, thereby directly impacting
the predicted rates of Pop III GRBs.

Resolving these theoretical uncertainties and conclusively identifying Pop III GRBs therefore requires
a much larger sample of high-redshift events. This goal is central to the design of missions like
the Einstein Probe (\emph{EP}; \citealt{2025SCPMA..6839501Y}) and the Space-based multi-band astronomical
Variable Objects Monitor (\emph{SVOM}; \citealt{2016arXiv161006892W}), which are optimized to detect
such events using highly sensitive instruments in the soft X-ray band---a strategy identified as
most effective by population studies \citep{2015MNRAS.448.2514G,2015JHEAp...7...35S,2025ApJ...988L..71W}.
The potential of this approach is already being demonstrated. Following their successful launches in 2024,
both \emph{EP} and \emph{SVOM} have reported significant high-redshift detections. \emph{EP}'s Wide-field
X-ray Telescope (WXT) has detected a GRB at $z=4.859$ (EP240315a; \citealt{2025NatAs...9..564L}), and
the coded-mask gamma-ray imager (ECLAIRs) on board \emph{SVOM} has captured GRB 250314A at $z=7.3$
\citep{2025arXiv250718783C}, which ranks it as the fifth-most distant GRB on record. These early
achievements not only validate the mission designs but also directly advance the core objective of
assembling the larger sample of high-redshift events needed to probe Pop III GRBs.

This work presents the first comprehensive, instrument-specific forecast for the detectability of
Pop III GRBs by the \emph{EP} and \emph{SVOM} missions. Our methodology follows the proven approach of
\cite{2017MNRAS.472.3532M}, applying their state-of-the-art $N$-body/hydrodynamical simulations to
compare Pop II/I and Pop III GRB redshift distributions and exploring different Pop III IMFs.
The primary aim is to bridge the gap between theory and upcoming observational capabilities.
We build two progenitor populations based on a critical metallicity threshold to quantify
the detectable Pop III GRB rates for \emph{EP}/WXT and \emph{SVOM}/ECLAIRs, thereby translating
theoretical rates into testable predictions. Our results establish a crucial observational benchmark
and outline a concrete strategy for identifying Pop III progenitors, marking a significant step
in preparing for the new era of time-domain astrophysics.

The rest of this paper is structured as follows. Section~\ref{sec:simulation} describes the hydrodynamical
chemistry simulations employed. Section~\ref{sec:method} outlines the method for calculating Pop III GRB
detection rates for detectors with specified energy bands and sensitivities. The resulting populations
of Pop III GRBs detectable by \emph{EP}/WXT and \emph{SVOM}/ECLAIRs are presented in Section~\ref{sec:result},
and the conclusions are given in Section~\ref{sec:conclusions}. Throughout this work, we adopt a standard
$\Lambda$CDM cosmological model with the following parameters: matter density $\Omega_{\rm m}=0.3$,
dark energy density $\Omega_{\Lambda}=0.7$, baryon density $\Omega_{b}=0.04$, reduced Hubble constant $h=0.7$,
amplitude of matter fluctuations $\sigma_8=0.9$, and primordial spectral index $n_s=1$.

\section{Numerical Simulations}
\label{sec:simulation}
This work employs the $N$-body/hydrodynamical simulations from \cite{2017MNRAS.472.3532M}, whose essential
features are summarized below. Further details are available in the original publication. The simulations
were run using a modified version of the \emph{\sc GADGET-2} code \citep{2005MNRAS.364.1105S} that
incorporates atomic and molecular non-equilibrium chemistry, resonant and fine-structure cooling, Pop III
and Pop II/I star formation with the corresponding IMFs, and metal pollution of various heavy elements
\citep{2007MNRAS.379..963M,2010MNRAS.407.1003M,2013MNRAS.435.1443M,2007MNRAS.382.1050T}. The simulations
span from $z=100$ to $z=5.5$, with $20$ snapshots saved between $z=17$ and $z=5.5$. The simulation box has
a side length of $\mathrm{10\,Mpc}$ $h^{-1}$ with a total of $2\times320^3$ particles, resulting in gas
and dark matter particle masses of $3.39\times10^5\,\mathrm{M_{\odot}}\,h^{-1}$ and $2.20\times10^6\,\mathrm{M_{\odot}}\,h^{-1}$,
respectively. Our model applies the chemical reaction network and associated molecular/metal cooling functions
from \cite{2007MNRAS.379..963M}. Star formation is triggered when a gas particle reaches a density of $\mathrm{70\,cm^{-3}}$
due to cooling. To prevent overcooling, these star-forming particles receive kinetic wind feedback of
$\mathrm{500\,km\, s^{-1}}$. We adopt stellar lifetimes from \cite{1993ApJ...416...26P}. Stars that end
their lives as SNe \citep{1993ApJ...416...26P,2010ApJ...724..341H} enrich their host galaxies
by yielding and dispersing heavy elements. The produced metals are distributed to neighboring particles
using the smoothed particle hydrodynamics (SPH) kernel, following the method of \cite{2007MNRAS.382.1050T}
to approximate the rapid mixing in the interstellar medium \citep{2002ApJ...581.1047D}.

In the $N$-body hydrodynamical chemistry simulations, the transition from primordial (Pop III) to Pop II/I
star formation is determined by the metallicity ($Z$) of the star-forming gas, which occurs once $Z$ exceeds
a critical value of $Z_{\rm crit}=10^{-4}\,Z_{\odot}$ \citep{2003Natur.425..812B,2003Natur.422..869S,2006MNRAS.369.1437S}.
This transition occurs rapidly, as pollution from the first stellar explosions boosts metallicities
to $\sim 10^{-3}\,Z_{\odot}$ within approximately $10^7$ years, making the exact value of the critical metallicity $Z_{\rm crit}$
a secondary concern \citep{2010MNRAS.407.1003M}. The dominant source of uncertainty stems from the assumed Pop III IMF.
To account for the ongoing uncertainty in primordial stellar properties, we explore three Pop III IMF scenarios:
very massive SNe (VMSN), massive SNe (MSN), and regular SNe (RSN) \citep{2017MNRAS.472.3532M,2017MNRAS.466.1140M}.

All models use a Salpeter IMF but differ in the stellar mass ranges and the corresponding mass ranges of
the SN progenitors that drive metal enrichment. The VMSN model considers very massive first stars of
$\mathrm{100\textrm{--}500\,M_\odot}$, with metal enrichment dominated by Pair-Instability Supernova
(PISN) progenitors in the range $\mathrm{140\textrm{--}260\,M_\odot}$ \citep{2002ApJ...567..532H}.
Both the MSN and RSN models assume the Pop III IMF over $\mathrm{[0.1,\,100]\,M_\odot}$,
differing in the mass ranges of the SN progenitors responsible for metal pollution: $\mathrm{[10,\,100]\,M_\odot}$
for the MSN model \citep{2010ApJ...724..341H} and $\mathrm{[10,\,40]\,M_\odot}$ for the RSN model
\citep{1995ApJS..101..181W, 2002ApJ...567..532H}. For Pop II/I stars, a Salpeter IMF over the mass range
$\mathrm{[0.1,\,100]\,M_\odot}$ is adopted, with metal yields from AGB stars \citep{1997A&AS..123..305V},
Type Ia SNe \citep{2003NuPhA.718..139T}, and Type II SNe \citep{1995ApJS..101..181W}. We assume an energy
release of $\mathrm{10^{51}\,erg}$ for all SNe except PISNe, for which the energy ranges from
$\sim \mathrm{10^{51}\textrm{--}10^{53}\,erg}$ based on progenitor mass \citep{2002ApJ...567..532H}.

\begin{figure}
\begin{center}
\includegraphics[width=0.48\textwidth]{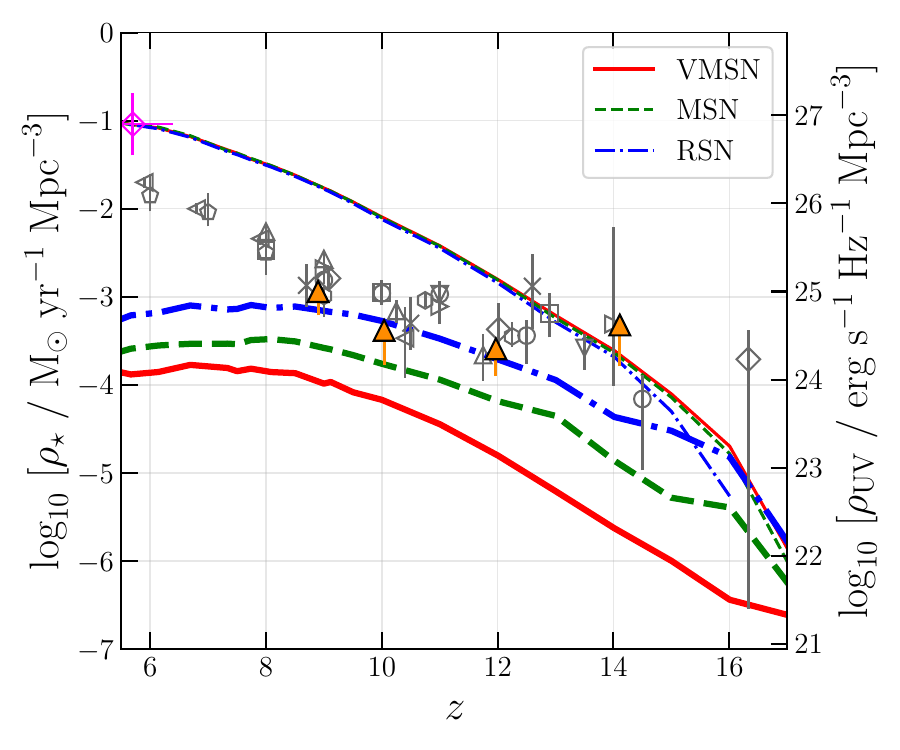}
\vskip-0.1in
\caption{Cosmic SFR density as a function of redshift for the VMSN (red solid lines), MSN (green dashed lines),
and RSN (blue dot-dashed lines) Pop III IMF models. Simulated SFR densities for Pop III and Pop II/I stars
are shown as thick and thin lines, respectively. Observational data include: a measurement from IR observations
(magenta diamond; \citealt{2016MNRAS.461.1100R}), spectroscopic lower limits from JWST UV data (orange triangles;
\citealt{2024ApJ...960...56H, 2025ApJ...980..138H}), and previous UV photometric estimates (gray points;
see \citealt{2025ApJ...980..138H} and references therein).}
\label{fig1}
\vskip-0.2in
\end{center}
\end{figure}

Figure~\ref{fig1} shows the comoving SFR density history, $\rho_\star(z)$, for Pop III
and Pop II/I stars from our three Pop III IMF models. The results are compared with observational estimates
from multi-wavelength observations (\citealt{2016MNRAS.461.1100R,2024ApJ...960...56H,2025ApJ...980..138H},
and references therein). While the simulated SFR densities for Pop II/I stars are very similar across
the models, the Pop III SFR densities differ significantly \citep{2010MNRAS.407.1003M,2017MNRAS.472.3532M}.
The simulated Pop II/I SFR densities are consistent with those from infrared (IR) observations at $z\sim6$
\citep{2016MNRAS.461.1100R} but are generally higher than most UV-based determinations at $5<z<12$,
including those from recent JWST data\footnote{The SFR density in these JWST studies is derived from
the measured UV luminosity density (right axis of Figure~\ref{fig1}) through the application of
a conversion factor, $1.15\times10^{-28}\,\mathrm{M_\odot\,yr^{-1}/(erg\,s^{-1}\,Hz^{-1})}$.}
\citep{2024ApJ...960...56H,2025ApJ...980..138H}. UV-based SFR estimates are potentially incomplete,
as they may underestimate embedded star formation by up to 1 dex
\citep{2012ARA&A..50..531K,2014ARA&A..52..415M,2016MNRAS.461.1100R}. Despite predictions of lower dust
content in early galaxies compared to $z<3$ objects \citep{2015Natur.522..455C}, fundamental questions
about high-$z$ IR luminosities and dust grain origin persist
\citep{2015MNRAS.451L..70M,2016A&A...587A.157B,2016ApJ...833...72B,2016MNRAS.463L.112F,2016ApJ...820...98L}.
Therefore, significant dust extinction may already be present at $z \sim 7$, hiding dusty, UV-faint galaxies
\citep{2013MNRAS.429.2718S,2016MNRAS.462.3130M,2016ApJ...823..128M}. Furthermore, high-resolution simulations
indicate that SFRs at $z > 10$ could be slightly underestimated \citep{2010MNRAS.407.1003M,2011MNRAS.416.2760C,2017MNRAS.472.3532M}.

\section{Analysis Method}
\label{sec:method}
We categorize the possible progenitors of long GRBs into two populations: (i) GRBII,
comprising GRBs that originate from Pop II/I stars, selected from star-forming particles with metallicity $Z$
exceeding a critical value of $Z_{\rm crit}=10^{-4}\,Z_{\odot}$; and (ii) GRBIII, comprising GRBs
from Pop III stars, selected from particles with $Z\leq Z_{\rm crit}$. The rapid pollution process, which boosts
metallicities to $\sim 10^{-3}\,Z_{\odot}$ within a few $10^7$ years after the first explosions,
makes uncertainties in $Z_{\rm crit}$ for the Pop III--Pop II/I transition of secondary importance
\citep{2010MNRAS.407.1003M}.

In the following, we describe the calculation of the GRB event rate (Section~\ref{subsec:Rate}) for
a detector with a specified energy band and sensitivity, for both Pop II/I and Pop III progenitors
(Sections~\ref{subsec:Pop II/I} and \ref{subsec:Pop III}).

\subsection{Event Rate of GRBs}
\label{subsec:Rate}
The differential number of GRBs detected per unit time, per redshift interval $\mathrm{d}z$, and
per luminosity interval $\mathrm{d}L$ is given by
\begin{equation}
\frac{\mathrm{d}^{3}N}{\mathrm{d}t\mathrm{d}z\mathrm{d}L}= \frac{\rho_{\rm GRB}(z)}{1+z}\frac{\mathrm{d}V(z)}{\mathrm{d}z}\Psi(L)\;,
\end{equation}
where $\rho_{\rm GRB}(z)$ is the comoving formation rate density of GRBs (in units of $\mathrm{Mpc^{-3}\;yr^{-1}}$),
the factor $(1+z)^{-1}$ accounts for cosmological time dilation, $\Psi(L)$ is the normalized GRB luminosity function
(LF), and $\mathrm{d}V(z)/\mathrm{d}z$ is the comoving volume element.

Within the context of the collapsar origin, the GRB formation rate density $\rho_{\mathrm{GRB},i}(z)$
for population $i$ is described as
\begin{equation}
\label{eq:rho_GRB}
\rho_{\mathrm{GRB},i}(z)=f_{\mathrm{GRB},i}\,\,\zeta_{\mathrm{BH},i}\,\,\rho_{\star,i}(z)\;,
\end{equation}
where $f_{\mathrm{GRB},i}$ is the fraction of stellar-mass BHs that produce a GRB, $\zeta_{\mathrm{BH},i}$
is the BH production efficiency per unit stellar mass (in  units of $\mathrm{M_\odot^{-1}}$), and $\rho_{\star,i}(z)$
is the comoving SFR density at redshift $z$ for population $i$ (in units of $\mathrm{M_\odot\,yr^{-1}\,Mpc^{-3}}$).

To compute the GRB event rate, we must specify the LF, $\Psi_{i}(L)$, for the two GRB populations considered here.
We refer the reader to \cite{2011MNRAS.416.2760C} for full details on the choice of $\Psi_{i}(L)$.
Specifically, for the GRBII population, we model the LF as a single power law with an exponential cutoff
at low luminosity (as derived in \citealt{2010MNRAS.407.1972C}):
\begin{equation}
\Psi(L)\propto \left(\frac{L}{L_{\rm cut}}\right)^{-\nu}\exp\left(-\frac{L_{\rm cut}}{L}\right)\;,
\end{equation}
where $\nu$ is the high-luminosity power-law index and $L_{\rm cut}$ is the cutoff luminosity. To account for
possible evolution in the GRB LF, the cutoff luminosity evolves with redshift as $L_{\rm cut}(z)=L_{\rm cut,0}(1+z)^{\delta}$,
where $L_{\rm cut,0}$ is its value at $z=0$. Given the similarity between the SFR evolution of our GRBII population
and the host galaxy sample in \citet{2010MNRAS.407.1972C}, we adopt the following parameters:
$L_{\rm cut,0}=0.3 \times 10^{50} \ \mathrm{erg \ s^{-1}}$, $\delta=2.0$, and $\nu=1.7$. This combination provides
a good fit to the existing observational data. For the GRBIII population, we assume that GRBs from Pop III
stars are significantly brighter, with typical luminosities expected to exceed $10^{53.6} \ \mathrm{erg \ s^{-1}}$
\citep{2011ApJ...731..127T}. We adopt a characteristic cutoff luminosity of $L_{\rm cut} = 10^{54} \ \mathrm{erg \ s^{-1}}$,
held constant with redshift, and $\nu\sim1.7$. We also explore the parameter ranges
$L_{\rm cut} = 10^{53} \text{--} 10^{55} \ \mathrm{erg \ s^{-1}}$ and $1.5 < \nu < 2.0$.
The dependence of our results on these parameters is discussed in Section~\ref{sec:result}.

For a detector operating in the $E_{1}\textrm{--}E_{2}$ energy band with a flux threshold of $P_{\rm lim}$,
the expected event rate of $\mathrm{GRB},i$ (in units of $\mathrm{yr^{-1}}$) at redshifts greater than $z$
can be calculated as
\begin{align}
\label{eq:RGRB}
\mathcal{R}_{\mathrm{GRB},i}(>z) = & \frac{\Omega}{4\pi} \, \gamma_{\mathrm{beam}} \, \eta_{\mathrm{duty}} \int_{z}^{\infty} \frac{\rho_{\mathrm{GRB},i}(z')}{1+z'} \, \frac{\mathrm{d}V(z')}{\mathrm{d}z'} \, \mathrm{d}z' \nonumber \\
& \times \int_{\max\left[L_{\mathrm{min}}, \, L_{\mathrm{lim}}(z')\right]}^{L_{\mathrm{max}}} \Psi_{i}(L') \, \mathrm{d}L'\,,
\end{align}
where $\Omega$ is the solid angle of the detector's field of view (FOV), $\gamma_{\rm beam}=5.5\times10^{-3}$
is the beaming factor of a relativistic jet with an average opening angle of $\sim6^{\circ}$
\citep{2007A&A...466..127G,2013MNRAS.428.1410G}, and $\eta_{\rm duty}$ is the duty cycle
(the fraction of total time spent on observations)\footnote{A significant fraction of
mission time is lost to satellite inactivation in the South Atlantic Anomaly as well as
to spacecraft slewing.}. The LF is normalized over the range $L_{\rm min}=10^{47}\,\mathrm{erg\,s^{-1}}$
to $L_{\rm max}=10^{57}\,\mathrm{erg\,s^{-1}}$. The luminosity threshold in Equation~(\ref{eq:RGRB})
is given by
\begin{equation}
\label{eq:Llim}
L_{\rm lim}(z)=4\pi D_{L}^{2}(z)P_{\rm lim}k(z)\;,
\end{equation}
where $D_{L}(z)$ is the luminosity distance and $k(z)$ is the spectral $k$-correction, which converts
the observed flux in the detector's energy band $[E_{1},\,E_{2}]$ to the rest-frame $1$--$10^{4}$ keV band.
The precise form of $k(z)$ depends on whether the detector's sensitivity limit, $P_{\rm lim}$,
is defined as a photon or energy flux. For a photon flux limit $P_{\rm lim}$ (in units of $\mathrm{\text{ph}\,cm^{-2}\,s^{-1}}$),
\begin{equation}
\label{eq:k1}
k(z)=\frac{\int_{1\,\mathrm{keV}/(1+z)}^{10^{4}\,\mathrm{keV}/(1+z)}E\,N(E)\mathrm{d}E}{\int_{E_{1}}^{E_{2}}N(E)\mathrm{d}E}\,.
\end{equation}
For an energy flux limit $P_{\rm lim}$
(in units of $\mathrm{erg\,cm^{-2}\,s^{-1}}$),
\begin{equation}
\label{eq:k2}
k(z)=\frac{\int_{1\,\mathrm{keV}/(1+z)}^{10^{4}\,\mathrm{keV}/(1+z)}E\,N(E)\mathrm{d}E}{\int_{E_{1}}^{E_{2}}E\,N(E)\mathrm{d}E}\,.
\end{equation}
The key distinction is in the denominator: it is the integrated photon flux for a photon flux limit,
but the integrated energy flux for an energy flux limit. In both cases, $N(E)$ is the observed photon
spectrum, which we model with a Band function \citep{Band1993ApJ}. We adopt typical values of
$-1$ and $-2.3$ for the low- and high-energy spectral indices, respectively
\citep{Kaneko2006ApJS,Nava2011A&A,2020ApJ...893...46V}. The spectral peak energy $E_{\rm peak}$
is determined from the luminosity $L$ via the empirical $E_{\rm peak}\textrm{--}L$ correlation
\citep{2004ApJ...609..935Y,Nava2012MNRAS}: $\log_{10} [E_{\rm peak} (1 + z)] = -25.33 + 0.53 \log_{10} L$.

\begin{table*}
\centering
\renewcommand\arraystretch{1.3}
\tabcolsep=0.3cm
\caption{GRB Production Fraction and BH Production Efficiency for Pop II/I and III GRBs
\hfill {\footnotesize (Calculated for various Pop III IMFs with $m_{\rm min} = 20\,\mathrm{M_\odot}$)}}
\begin{tabular}{lcccccc}
\hline
\hline
 & \multicolumn{2}{c}{Pop II/I $\mathrm{GRB}^{\textsuperscript{a}}$}  &   & \multicolumn{3}{c}{Pop III GRB} \\
\cline{2-3} \cline{5-7}
Model  & $f_{\mathrm{GRBII}}$  & $\zeta_{\mathrm{BHII}}$ ($\mathrm{M_\odot^{-1}}$) &   & $f_{\mathrm{GRBIII_{up1}}}$ & $f_{\mathrm{GRBIII_{up2}}}$  & $\zeta_{\mathrm{BHIII}}$ ($\mathrm{M_\odot^{-1}}$)\\
\hline
VMSN  & $1.0\times10^{-3}$  & $2.0\times10^{-3}$ &   & ${1.3\times10^{-3}}^{\textsuperscript{b}}$ & ${2.6\times10^{-3}}^{\textsuperscript{b}}$  & ${3.2\times10^{-3}}^{\textsuperscript{b}}$\\

MSN  & $1.0\times10^{-3}$  & $2.0\times10^{-3}$ &   & $1.0\times10^{-3}$ & $2.0\times10^{-3}$  & $2.0\times10^{-3}$\\

RSN  & $1.1\times10^{-3}$  & $2.0\times10^{-3}$ &   & $3.8\times10^{-4}$ & $7.6\times10^{-4}$  & $2.0\times10^{-3}$\\
\hline
\end{tabular}
\label{tab1}
\\
\noindent{\footnotesize{\textsuperscript{a} For Pop II/I stars, a Salpeter IMF over the mass range
$\mathrm{[0.1,\,100]\,M_\odot}$ is adopted.}}\\
\noindent{\footnotesize{\textsuperscript{b} Unlike the MSN and RSN models, the inferred values in the VMSN model are independent of $m_{\rm min}$.}}
\end{table*}

\begin{table*}
\centering
\renewcommand\arraystretch{1.3}
\tabcolsep=0.3cm
\caption{GRB Production Fraction and BH Production Efficiency for Pop II/I and III GRBs
\hfill {\footnotesize (Calculated for various Pop III IMFs with $m_{\rm min} = 40\,\mathrm{M_\odot}$)}}
\begin{tabular}{lcccccc}
\hline
\hline
 & \multicolumn{2}{c}{Pop II/I $\mathrm{GRB}^{\textsuperscript{a}}$}  &   & \multicolumn{3}{c}{Pop III GRB} \\
\cline{2-3} \cline{5-7}
Model  & $f_{\mathrm{GRBII}}$  & $\zeta_{\mathrm{BHII}}$ ($\mathrm{M_\odot^{-1}}$) &   & $f_{\mathrm{GRBIII_{up1}}}$ & $f_{\mathrm{GRBIII_{up2}}}$  & $\zeta_{\mathrm{BHIII}}$ ($\mathrm{M_\odot^{-1}}$)\\
\hline
VMSN  & $3.2\times10^{-3}$  & $6.2\times10^{-4}$ &   & ${1.3\times10^{-3}}^{\textsuperscript{b}}$ & ${2.6\times10^{-3}}^{\textsuperscript{b}}$  & ${3.2\times10^{-3}}^{\textsuperscript{b}}$\\

MSN  & $3.2\times10^{-3}$  & $6.2\times10^{-4}$ &   & $3.2\times10^{-3}$ & $6.4\times10^{-3}$  & $6.2\times10^{-4}$\\

RSN  & $3.5\times10^{-3}$  & $6.2\times10^{-4}$ &   & $1.2\times10^{-3}$ & $2.4\times10^{-3}$  & $6.2\times10^{-4}$\\
\hline
\end{tabular}
\label{tab2}
\\
\noindent{\footnotesize{\textsuperscript{a} For Pop II/I stars, a Salpeter IMF over the mass range
$\mathrm{[0.1,\,100]\,M_\odot}$ is adopted.}}\\
\noindent{\footnotesize{\textsuperscript{b} Unlike the MSN and RSN models, the inferred values in the VMSN model are independent of $m_{\rm min}$.}}
\end{table*}

\subsection{GRBs from Pop II/I Stars}
\label{subsec:Pop II/I}
The GRB formation rate density (see Equation~(\ref{eq:rho_GRB})) can be estimated from the SFR density
of population $i$, given the parameters $f_{\mathrm{GRB},i}$ and $\zeta_{\mathrm{BH},i}$.

For Pop II/I stars, the BH production efficiency per unit stellar mass is determined by
\begin{equation}
\label{eq:BH}
\zeta_{\mathrm{BHII}}=\frac{\int_{m_{\rm min}}^{100}
\phi(m_{\star})\mathrm{d}m_{\star}}{\int_{0.1}^{100}m_{\star}\phi(m_{\star})\mathrm{d}m_{\star}}\,,
\end{equation}
where $m_{\rm min}$ is the minimum stellar mass of stars that form BHs and
\begin{equation}
\phi(m_{\star})\propto m_{\star}^{-2.3}
\end{equation}
is the adopted Salpeter IMF \citep{1955ApJ...121..161S}.
Given the uncertainties in $m_{\rm min}$, we consider two cases with $m_{\rm min}=20\,\mathrm{M_\odot}$
and $m_{\rm min}=40\,\mathrm{M_\odot}$, following the approach of \cite{2011MNRAS.416.2760C}.
For $m_{\rm min}=20\,\mathrm{M_\odot}$ ($40\,\mathrm{M_\odot}$), we find
$\zeta_{\mathrm{BHII}}=2.0\times10^{-3}\,\mathrm{M_\odot^{-1}}$ ($0.62\times10^{-3}\,\mathrm{M_\odot^{-1}}$).

Since not all BHs produce long GRBs, we have to account for the fraction via the parameter $f_{\mathrm{GRB},i}$
in Equation~(\ref{eq:rho_GRB}). We calibrate this parameter using GRBs detected by the Burst Alert Telescope
(BAT) on board the \emph{Swift} satellite. Adopting a photon flux limit of $P_{\rm lim}=0.4\,\mathrm{ph\,cm^{-2}\,s^{-1}}$
(15--150 keV), we select 1403 bursts with 1-s peak flux $P\ge0.4\,\mathrm{ph\,cm^{-2}\,s^{-1}}$ from
a parent sample of 1467 long-duration ($T_{90}\ge2\,\mathrm{s}$) GRBs observed by \emph{Swift} up to October 1,
2025.\footnote{\url{https://swift.gsfc.nasa.gov/archive/grb_table/}} As the simulations stop at $z=5.5$,
we calibrate $f_{\mathrm{GRB},i}$ against the \emph{Swift} GRB rate at $z>5.5$. Previous studies indicate that $\sim2\%$
of the entire \emph{Swift} sample lies at $z>5.5$ \citep{2016ApJ...817....7P}, implying that \emph{Swift} has detected
approximately 28 GRBs at these redshifts. Considering \emph{Swift}/BAT's FOV ($\mathrm{1.4\,sr}$),
a mission duration of $\sim20.8$ years, and an average duty cycle of 78\% \citep{2016ApJ...829....7L},
the detection rate for GRBs at $z>5.5$ with $P\ge0.4$ $\mathrm{ph\,cm^{-2}\,s^{-1}}$ (15--150 keV)
is estimated to be $\sim1.2$ $\mathrm{events\,yr^{-1}\,sr^{-1}}$. Using this observed rate
as a constraint and assuming $m_{\rm min}=20\,\mathrm{M_\odot}$, we find $f_{\mathrm{GRBII}}=1.0\times10^{-3}$,
$1.0\times10^{-3}$, and $1.1\times10^{-3}$ for the VMSN, MSN, and RSN Pop III IMF models, respectively.
Repeating the calculation for $m_{\rm min}=40\,\mathrm{M_\odot}$ yields higher $f_{\mathrm{GRBII}}$ values
of $3.2\times10^{-3}$, $3.2\times10^{-3}$, and $3.5\times10^{-3}$ for the same model sequence.
These results for $m_{\rm min}=20\,\mathrm{M_\odot}$ and $m_{\rm min}=40\,\mathrm{M_\odot}$
are summarized in Table~\ref{tab1} and Table~\ref{tab2}, respectively.

\subsection{GRBs from Pop III Stars}
\label{subsec:Pop III}
Similarly, we must specify the parameters $f_{\mathrm{GRBIII}}$ and $\zeta_{\mathrm{BHIII}}$ for
the GRBIII population.

In the VMSN model, the first stars form over a wide mass range of $100$--$500\,\mathrm{M_\odot}$.
However, those with initial masses between $140$ and $260\,\mathrm{M_\odot}$ are predicted to
explode as PISNe \citep{1971reas.book.....Z,1999sar..book.....Z,2002ApJ...567..532H},
which completely disrupt the star and leave no compact remnant. Therefore, BHs, and hence
the GRBs they power, can only form from progenitors in the $100$--$140\,\mathrm{M_\odot}$ and
$260$--$500\,\mathrm{M_\odot}$ intervals. For the adopted VMSN Pop III IMF, this yields a BH
production efficiency of $\zeta_{\mathrm{BHIII}}=3.2\times10^{-3}\,\mathrm{M_\odot^{-1}}$,
calculated similarly to Equation~(\ref{eq:BH}). For the MSN and RSN Pop III models,
which use the same stellar mass range as the Pop II/I IMF ($\mathrm{0.1\textrm{--}100\,M_\odot}$),
we apply the same BH production efficiency as for Pop II/I stars. This gives
$\zeta_{\mathrm{BHIII}}=2.0\times10^{-3}\,\mathrm{M_\odot^{-1}}$ for $m_{\rm min}=20\,\mathrm{M_\odot}$
(Table~\ref{tab1}) and $\zeta_{\mathrm{BHIII}}=0.62\times10^{-3}\,\mathrm{M_\odot^{-1}}$
for $m_{\rm min}=40\,\mathrm{M_\odot}$ (Table~\ref{tab2}).

The fraction of Pop III stars producing GRBs is currently unconstrained. To date, there is no
conclusive evidence for a Pop III GRB in the \emph{Swift} catalog or any other archive. Even
the most distant GRB known exhibits prompt and afterglow properties consistent with the low-redshift
Pop II GRB \citep{2009Natur.461.1258S}, suggesting a non-Pop III origin. Given the non-detection
of any confirmed Pop III GRB by \emph{Swift}, we can place a firm upper limit on the Pop III GRB
production fraction:
\begin{equation}
\frac{\mathcal{R}_{\mathrm{GRBIII}}(z>5.5)}{\mathcal{R}_{\mathrm{Swift}}(z>5.5)}<\frac{1}{N_{\mathrm{Swift}}}\,,
\end{equation}
where $\mathcal{R}_{\mathrm{Swift}}(z>5.5)=\mathrm{1.2\,events\,yr^{-1}\,sr^{-1}}$ is the \emph{Swift}
GRB rate at $z>5.5$ and $N_{\mathrm{Swift}}$ is the number of GRBs at $z>5.5$ detected by \emph{Swift}.
We derive two upper limits for $f_{\mathrm{GRBIII}}$ based on different sample criteria (Tables~\ref{tab1} and~\ref{tab2}).
The first limit is derived from the entire estimated sample of $\sim$28 GRBs at $z>5.5$ (inferred from
the finding that $\sim 2\%$ of the \emph{Swift} sample occurs at $z>5.5$; \citealt{2016ApJ...817....7P}),
yielding $f_{\mathrm{GRBIII_{up1}}}< 1.3\times10^{-3}$ for the VMSN model. However, since incomplete
follow-up observations may have missed some high-redshift events, a Pop III GRB could be hidden among
\emph{Swift} bursts without redshift measurement. To account for this, we calculate a second, less
stringent limit, $f_{\mathrm{GRBIII_{up2}}}$, based solely on the 14 bursts with confirmed redshifts
$z>5.5$, giving $f_{\mathrm{GRBIII_{up2}}}< 2.6\times10^{-3}$ for the VMSN model. Note that
for the MSN and RSN models, the inferred upper limits of $f_{\mathrm{GRBIII_{up1}}}$ and $f_{\mathrm{GRBIII_{up2}}}$
depend on the assumed minimum BH progenitor mass $m_{\rm min}$.
The resulting values are as follows:
\begin{itemize}
  \item For $m_{\rm min}=20\,\mathrm{M_\odot}$ (Table~\ref{tab1}):

     MSN: $f_{\mathrm{GRBIII_{up1}}}< 1.0\times10^{-3}$, $f_{\mathrm{GRBIII_{up2}}}< 2.0\times10^{-3}$

     RSN: $f_{\mathrm{GRBIII_{up1}}}< 3.8\times10^{-4}$, $f_{\mathrm{GRBIII_{up2}}}<7.6\times10^{-4}$;
  \item For $m_{\rm min}=40\,\mathrm{M_\odot}$ (Table~\ref{tab2}):

     MSN: $f_{\mathrm{GRBIII_{up1}}}< 3.2\times10^{-3}$, $f_{\mathrm{GRBIII_{up2}}}< 6.4\times10^{-3}$

     RSN: $f_{\mathrm{GRBIII_{up1}}}< 1.2\times10^{-3}$, $f_{\mathrm{GRBIII_{up2}}}< 2.4\times10^{-3}$.
\end{itemize}

\section{Pop III GRBs Accessible by EP and SVOM}
\label{sec:result}
\begin{figure*}
\begin{center}
\includegraphics[width=0.48\textwidth]{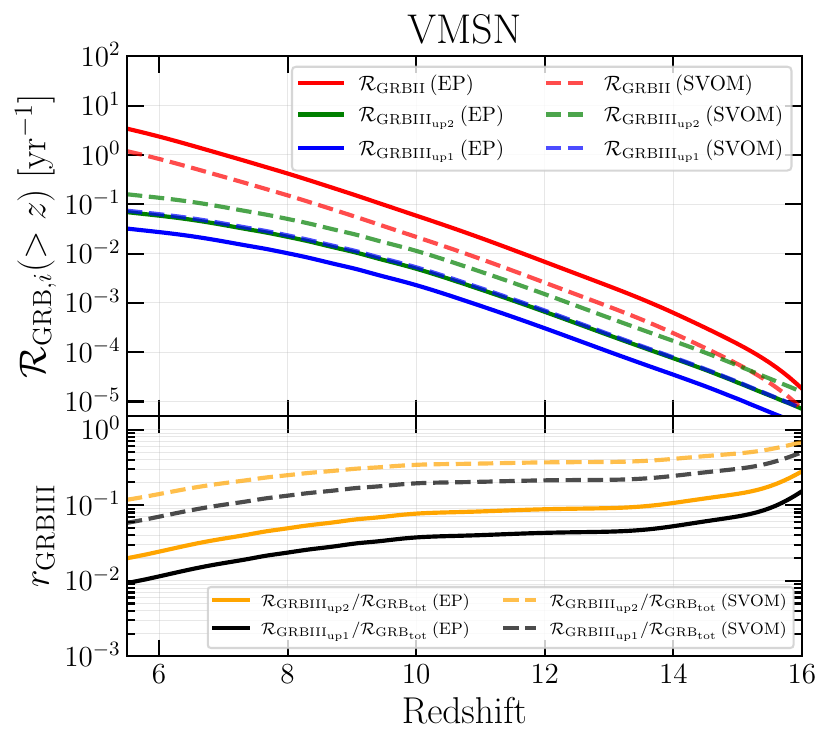}
\includegraphics[width=0.48\textwidth]{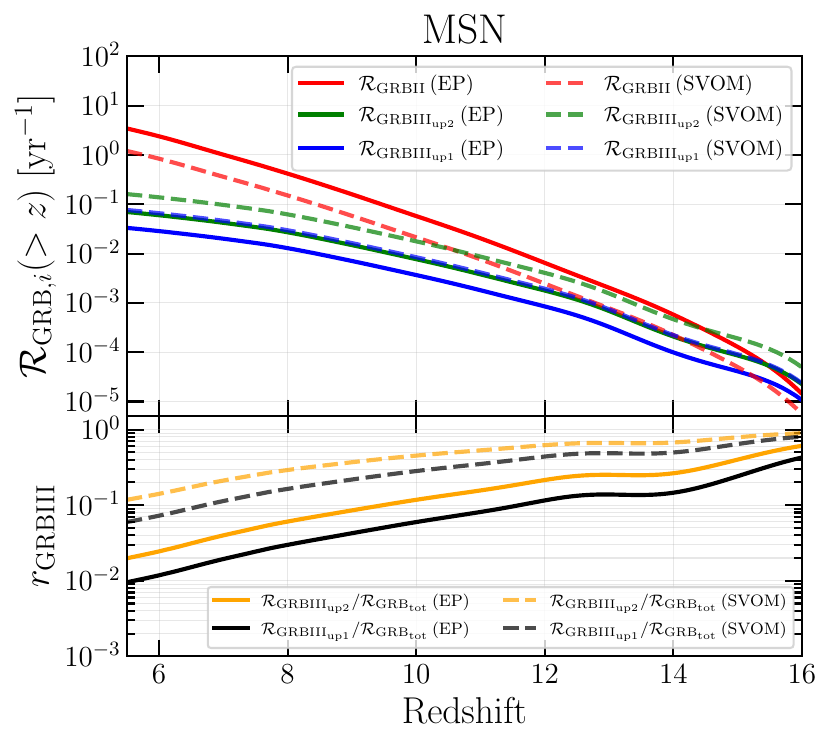}
\includegraphics[width=0.48\textwidth]{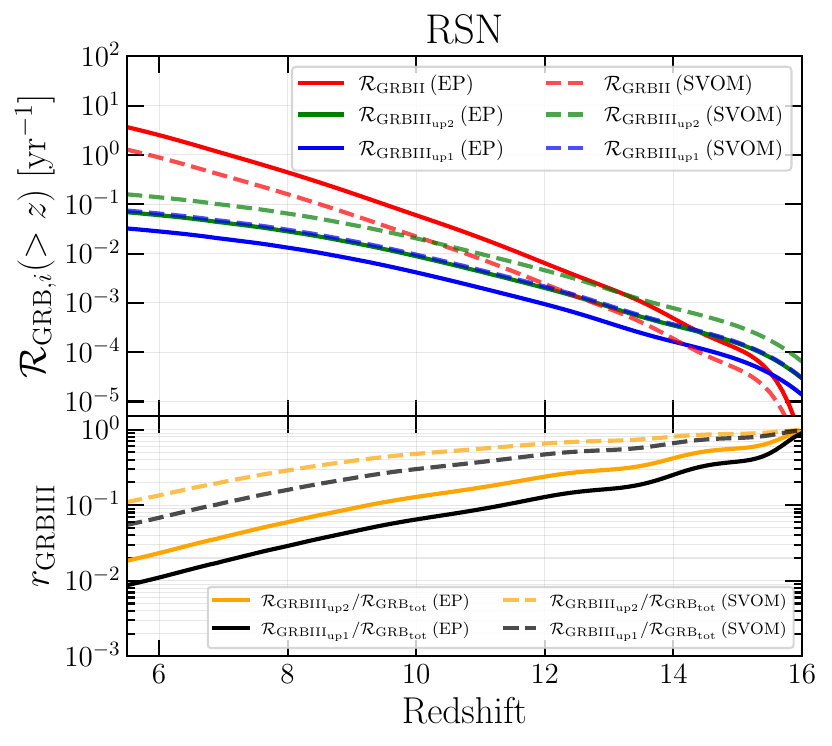}
\caption{Top panels: cumulative observed rate versus redshift for different GRB populations under
various Pop III IMF models. Red lines show the rates of GRBII detected with \emph{EP}/WXT (solid)
and \emph{SVOM}/ECLAIRs (dashed). Green and blue lines show the corresponding upper limits on
GRBIII rates derived from $f_{\mathrm{GRBIII_{up2}}}$ and $f_{\mathrm{GRBIII_{up1}}}$, respectively.
Bottom panels: redshift evolution of the GRBIII fraction,
$\mathcal{R}_{\mathrm{GRBIII}_{\mathrm{up},i}}/(\mathcal{R}_{\mathrm{GRBII}}+\mathcal{R}_{\mathrm{GRBIII}_{\mathrm{up},i}})$,
for various Pop III IMF models. Orange and black lines show the ratios for \emph{EP}/WXT (solid) and
\emph{SVOM}/ECLAIRs (dashed), calculated with $f_{\mathrm{GRBIII_{up2}}}$ and $f_{\mathrm{GRBIII_{up1}}}$,
respectively.}
\label{fig2}
\vskip-0.2in
\end{center}
\end{figure*}

Using the calibrated parameters $f_{\mathrm{GRB},i}$ and $\zeta_{\mathrm{BH},i}$,
we can predict the GRB detection rate for an instrument in a specific energy band at a given flux
threshold by integrating Equation~(\ref{eq:RGRB}). Figure~\ref{fig2} displays the resulting redshift
distributions of detected GRBs from different populations, as expected from observations by
\emph{EP}/WXT and \emph{SVOM}/ECLAIRs (solid lines for \emph{EP}/WXT and dashed ones for \emph{SVOM}/ECLAIRs).
It is important to note that while the choice of minimum BH progenitor mass $m_{\rm min}$
might affect the BH production efficiency $\zeta_{\mathrm{BH},i}$, it does not alter the redshift evolution
of GRBs or the predicted GRB rates. This invariance arises because the GRB production fraction
$f_{\mathrm{GRB},i}$ is calibrated against the fixed \emph{Swift} rate, ensuring the product
$f_{\mathrm{GRB},i}\cdot\zeta_{\mathrm{BH},i}$ remains constant.

\begin{table*}
\renewcommand\arraystretch{1.3}
\tabcolsep=0.3cm
\centering \caption{Expected Fractions of Pop III GRBs for \emph{EP}/WXT and \emph{SVOM}/ECLAIRs
across Pop III IMF models}
\begin{tabular}{lccccccccc}
\hline
\hline
 & \multicolumn{9}{c}{Pop III GRB Fraction Calculated Using $f_{\mathrm{GRBIII_{up2}}}$}   \\
 \cline{2-10}
Model & \multicolumn{4}{c}{\emph{EP}/WXT}  &   & \multicolumn{4}{c}{\emph{SVOM}/ECLAIRs} \\
\cline{2-5} \cline{7-10}
  & $z>6$ & $z>10$  & $z>14$ & $z>16$ &   & $z>6$ & $z>10$  & $z>14$ & $z>16$ \\
\hline
VMSN  & $\le2\%$  & $\le8\%$  & $\le11\%$ & $\le28\%$ &   & $\le14\%$  & $\le34\%$ & $\le41\%$ & $\le68\%$   \\

MSN   & $\le2\%$  & $\le12\%$  & $\le26\%$ & $\le61\%$  &   & $\le14\%$  & $\le45\%$ & $\le68\%$ & $\le90\%$   \\

RSN   & $\le2\%$  & $\le13\%$  & $\le42\%$ & $\le95\%$  &   & $\le14\%$  & $\le48\%$ & $\le81\%$ & $\le99\%$   \\
\hline
\end{tabular}
\label{tab3}
\end{table*}

In our calculation, we use \emph{EP}/WXT with an FOV of $\mathrm{1.1\,sr}$, a duty cycle of 67\%, and
a 10-s exposure sensitivity of $8.9\times 10^{-10}$ $\mathrm{\rm erg\,cm^{-2}\,s^{-1}}$ in the 0.5--4 keV
energy range \citep{2025SCPMA..6839501Y}. Similarly, for \emph{SVOM}/ECLAIRs, we use an FOV of $\mathrm{2.0\,sr}$,
an 85\% duty cycle, and a sensitivity of $7.2\times 10^{-8}$ $\mathrm{\rm erg\,cm^{-2}\,s^{-1}}$ over 4--150 keV
for a 10-s exposure \citep{2016arXiv161006892W}. The top panel of each plot in Figure~\ref{fig2} shows
the cumulative observed rate versus redshift for different GRB populations under various Pop III IMF models.
Red lines represent the rates of GRBII ($\mathcal{R}_{\mathrm{GRBII}}$) observable by \emph{EP}/WXT (solid)
and \emph{SVOM}/ECLAIRs (dashed), respectively. Green and blue lines represent the corresponding upper limits
for GRBIII rates ($\mathcal{R}_{\mathrm{GRBIII}_{\mathrm{up},i}}$), calculated using $f_{\mathrm{GRBIII_{up2}}}$
and $f_{\mathrm{GRBIII_{up1}}}$, respectively. The bottom panel shows the redshift evolution of the fraction of
GRBs originating from Pop III stars for various Pop III IMF models. This fraction is defined as
$r_\mathrm{GRBIII}=\mathcal{R}_{\mathrm{GRBIII}_{\mathrm{up},i}}/\mathcal{R}_{\mathrm{GRB_{tot}}}$, where
$\mathcal{R}_{\mathrm{GRB_{tot}}}=\mathcal{R}_{\mathrm{GRBII}}+\mathcal{R}_{\mathrm{GRBIII}_{\mathrm{up},i}}$
is the total GRB rate from both Pop II/I and Pop III GRBs. Here, solid and dashed orange lines correspond to
the fractions calculated with $f_{\mathrm{GRBIII_{up2}}}$ for \emph{EP}/WXT and \emph{SVOM}/ECLAIRs,
while black lines show the equivalent fractions computed using $f_{\mathrm{GRBIII_{up1}}}$.

In the VMSN model, considering the total GRB rate, \emph{EP}/WXT is expected to detect $\sim2.4$ GRBs
per year at $z>6$, compared to $\sim0.9$ for \emph{SVOM}/ECLAIRs. This higher detection rate
stems from \emph{EP}/WXT's softer energy band and enhanced sensitivity, which provide a greater
capability for detecting high-redshift ($z\ge6$) GRBs. These results are consistent with previous
theoretical predictions \citep{2015MNRAS.448.2514G,2015JHEAp...7...35S,2024ApJ...976L..16M,2025ApJ...988L..71W}.
The upper limits for the Pop III GRB rates at $z>6$ are below $0.03$ and $0.06\,\mathrm{events\,yr^{-1}}$
for \emph{EP}/WXT and \emph{SVOM}/ECLAIRs, respectively, when calculated with $f_{\mathrm{GRBIII_{up1}}}$.
The less stringent upper limits derived from $f_{\mathrm{GRBIII_{up2}}}$ are below $0.06$ and
$0.13\,\mathrm{events\,yr^{-1}}$, respectively. Furthermore, the expected fraction of Pop III GRBs
shows a clear increasing trend with redshift for both instruments. Using the $f_{\mathrm{GRBIII_{up2}}}$
upper limit, the detectable GRBIII fraction for \emph{EP}/WXT (\emph{SVOM}/ECLAIRs) is $\le2\%$ ($\le14\%$)
at $z>6$, rising to $\le8\%$ ($\le34\%$) at $z>10$, $\le11\%$ ($\le41\%$) at $z>14$, and $\le28\%$ ($\le68\%$)
at $z>16$ (see Table~\ref{tab3}).
For the more restrictive $f_{\mathrm{GRBIII_{up1}}}$ case, these fractions are approximately a factor of two lower.
It is often hypothesized that Pop III GRBs could be significantly brighter than those from Pop II/I stars
\citep{2011ApJ...731..127T}. This is potentially important because the peak energy of a GRB is strongly
correlated with its luminosity \citep{2004ApJ...609..935Y}. Consequently, the harder energy band of
\emph{SVOM}/ECLAIRs, compared to that of \emph{EP}/WXT, makes it more sensitive to such potentially
luminous Pop III GRBs, resulting in a higher detectable fraction.

In the MSN and RSN models, similar trends are found, though with higher detectable GRBIII fractions
for both instruments (see Figure~\ref{fig2} and Table~\ref{tab3}). This systematic increase aligns with
the progression of the overall Pop III SFR density, which rises across the VMSN, MSN, and RSN models,
as shown in Figure~\ref{fig1}. As expected, the contribution of Pop III GRBs increases with redshift,
becoming dominant at $z>16$ for all Pop III IMF models considered. Therefore, any GRB detected at
such extreme redshifts would most likely be of Pop III origin, consistent with previous findings \citep{2011MNRAS.416.2760C}.
Our work builds upon the foundational studies of \cite{2017MNRAS.472.3532M}, and while our results
are consistent with the broader picture from \cite{2011MNRAS.416.2760C}, this study's contribution
is novel and timely. Specifically, this is the first work to deliver concrete, instrument-specific
predictions for Pop III GRB detectability with \emph{EP} and \emph{SVOM}. By focusing on the unique
soft X-ray capabilities of \emph{EP}/WXT and \emph{SVOM}/ECLAIRs, we provide the community with
a clear set of expectations and a strategy for identifying these elusive events.

Note that our results exhibit little dependence on the assumed LF for Pop III GRBs. The predicted
Pop III GRB rate at $z>6$ changes by less than a factor of two for a characteristic cutoff luminosity
$L_{\rm cut}$ between $10^{53}$ and $10^{55}\,\mathrm{erg\,s^{-1}}$ and for a slope $\nu$ in the range
of $1.5 < \nu < 2.0$.

\section{Summary and Discussions}\label{sec:conclusions}
The ongoing detection of long-duration GRBs at extremely high redshifts underscores
their profound potential as probes of the early Universe. Theoretical studies suggest
that massive, metal-free Pop III stars could produce exceptionally luminous GRBs,
with detectability potentially extending beyond $z>20$ \citep{2006ApJ...642..382B,
2010ApJ...715..967M,2011ApJ...726..107S,2011ApJ...731..127T}. Consequently,
identifying Pop III GRBs offers a direct avenue to investigate the properties of
the first stars \citep{2000ApJ...536....1L,2002ApJ...575..111B,2011MNRAS.416.2760C,
2011MNRAS.414..847S,2012ApJ...760...27W,2015NewAR..67....1W,2015JHEAp...7...35S,2016SSRv..202..159T}.
Therefore, quantifying the contribution of Pop III stars to early star formation
and their resultant GRBs represents a critical frontier in observational cosmology.
This effort provides a direct empirical test of the $\Lambda$CDM paradigm,
probing key predictions for the timeline of early structure assembly
(e.g., \citealt{2023ARA&A..61...65K}), the IMF of primordial stars (e.g.,
\citealt{2022MNRAS.511.2505L}), and the feedback processes that drove cosmic
reionization (e.g., \citealt{2023ARA&A..61...65K}).

In this work, we evaluate the prospects for detecting Pop III GRBs with \emph{EP}/WXT and
\emph{SVOM}/ECLAIRs. To model the expected redshift distributions and event rates of Pop
III and Pop II/I GRBs, we employ a series of $N$-body/hydrodynamical cosmological simulations from \cite{2017MNRAS.472.3532M}.
Within the collapsar scenario, we classify GRB progenitors into two metallicity-based categories:
GRBII, originating from Pop II/I stars with $Z>10^{-4}\,Z_{\odot}$, and GRBIII, from Pop III stars
with $Z\le10^{-4}\,Z_{\odot}$. To account for the persistent uncertainty in the characteristic
mass of Pop III stars, we consider three distinct Pop III IMF models: (i) a VMSN model with
stellar masses in $\mathrm{[100,\,500]\,M_\odot}$, where metal enrichment is driven by
PISN from progenitors of $\mathrm{[140,\,260]\,M_\odot}$; (ii) an MSN model spanning
$\mathrm{[0.1,\,100]\,M_\odot}$, enriched by SNe from $\mathrm{[10,\,100]\,M_\odot}$ progenitors;
and (iii) an RSN model, identical to the MSN case but with enrichment restricted to SNe from
$\mathrm{[10,\,40]\,M_\odot}$ progenitors. Finally, using a population synthesis framework
calibrated to \emph{Swift} GRB observations, we compute the detection rates of Pop III
and Pop II/I GRBs observable by \emph{EP}/WXT and \emph{SVOM}/ECLAIRs.

Within the VMSN model, the predicted annual detection rates of Pop II/I GRBs at $z>6$ are $\sim2.4$
for \emph{EP}/WXT and $\sim0.9$ for \emph{SVOM}/ECLAIRs, consistent with recent theoretical
predictions \citep{2025ApJ...988L..71W}. This aligns with the broader literature.
For instance, for the flux threshold and energy band of \emph{Swift}/BAT, \cite{2011MNRAS.416.2760C}
estimated a comparable Pop II/I GRB detection rate of $\sim1.0\,\mathrm{events\,yr^{-1}}$
at $z>6$, a value supported by both observational constraints (e.g.,
\citealt{2009AJ....138.1690P,2011A&A...526A..30G,2014ApJ...781....1L,
2022A&A...665A.125R,2025A&A...695A.239B}) and theoretical studies (e.g.,
\citealt{2009MNRAS.396..299S,2010ApJ...711..495B,2010MNRAS.406.1944W,2021A&A...649A.166P,2022ApJ...932...10G}).

To constrain the contribution from Pop III GRBs, we derive an upper limit on the fraction of
Pop III stars capable of producing a GRB, based on the assumption that none of the \emph{Swift}-detected
bursts with known redshifts $z>5.5$ originated from a Pop III progenitor. This constraint yields
upper limits of less than $0.06$ and $0.13\,\mathrm{events\,yr^{-1}}$ on the detectable rate of
Pop III GRBs at $z>6$ for \emph{EP}/WXT and \emph{SVOM}/ECLAIRs, respectively. Notably,
our constraints are consistent with earlier theoretical work: \cite{2011MNRAS.416.2760C} derived
an upper limit of fewer than $0.08\,\mathrm{events\,yr^{-1}}$ for Pop III GRB detections at $z>6$
with \emph{Swift}/BAT, a value that aligns closely with our own. More recently, \cite{2025arXiv250803689M}
estimated a detectable rate of $\sim0.22$ Pop III GRBs per year for a \emph{Swift}/BAT-like instrument.
Although this rate lies above the earlier upper limits, it is still compatible with a class of
more optimistic theoretical models \citep{2011MNRAS.416.2760C}. In a further comparison,
\cite{2019ApJ...878..128K} suggested that if more than 10\% of Pop III stars produce
an ultra-long GRB, future missions such as \emph{HiZ-GUNDAM} and \emph{THESEUS} would
detect up to one such event per year. This predicted detection rate remains considerably
higher than our estimates, reflecting differences in underlying model assumptions.

The GRB production fraction $f_{\mathrm{GRB},i}$ in our framework is determined by
the BH production efficiency $\zeta_{\mathrm{BH},i}$, which is calibrated via the Salpeter IMF.
To investigate the impact of the IMF, we perform parallel calculations using
the Kroupa \citep{2001MNRAS.322..231K} and Chabrier \citep{2003PASP..115..763C} IMFs.
The Kroupa IMF is a broken power law:
\begin{equation}
    \phi(m_{\star}) \propto\left\lbrace \begin{array}{lll}m_{\star}^{-0.3}; ~~~~~~~m_{\star}<0.08\,\mathrm{M_\odot} \\
                                          m_{\star}^{-1.3}; ~~~~~~~0.08\,\mathrm{M_\odot}\le m_{\star}<0.50\,\mathrm{M_\odot} \\
                                          m_{\star}^{-2.3}; ~~~~~~~m_{\star}\ge0.50\,\mathrm{M_\odot}\;. \\
    \end{array} \right.
\end{equation}
The Chabrier IMF uses a lognormal form for low masses and a power law for high masses:
\begin{equation}
    \phi(m_{\star}) \propto\left\lbrace \begin{array}{ll}\frac{1}{m_{\star}}\exp[-\frac{\left(\log_{10}m_{\star}-\log_{10}m_{c}\right)^2}{2\sigma^2}]; ~~~m_{\star}\le1.0\,\mathrm{M_\odot} \\
                                          m_{\star}^{-2.3}; ~~~~~~~~~~~~~~~~~~~~~~~~~~~~~~~~~~~m_{\star}>1.0\,\mathrm{M_\odot}\;, \\
    \end{array} \right.
\end{equation}
with $m_{c}=0.079\,\mathrm{M_\odot}$ and $\sigma=0.69$.

The derived values of $\zeta_{\mathrm{BHII}}$ for a minimum progenitor mass of
$m_{\rm min}=20\,\mathrm{M_\odot}$ are $3.6\times10^{-3}\,\mathrm{M_\odot^{-1}}$ (Kroupa) and
$4.6\times10^{-3}\,\mathrm{M_\odot^{-1}}$ (Chabrier), compared to our fiducial Salpeter value
of $2.0\times10^{-3}\,\mathrm{M_\odot^{-1}}$. For $m_{\rm min}=40\,\mathrm{M_\odot}$,
the corresponding values are $1.1\times10^{-3}\,\mathrm{M_\odot^{-1}}$ (Kroupa),
$1.5\times10^{-3}\,\mathrm{M_\odot^{-1}}$ (Chabrier), and $0.62\times10^{-3}\,\mathrm{M_\odot^{-1}}$
(Salpeter). In both cases, the choice of IMF introduces an uncertainty of approximately
a factor of $\sim2.5$ in $\zeta_{\mathrm{BHII}}$. As this variation does not alter
the conclusions of our rate forecasts, and to maintain consistency with the foundational
simulations of \cite{2017MNRAS.472.3532M} upon which our population synthesis is built,
we adopt the Salpeter IMF for all results presented in this work.

While Pop III GRBs are found to be subdominant to normal Pop II/I events at $z<10$,
their fractional contribution rises significantly with redshift. At $z>10$, the predicted
Pop III fraction is $\sim8\%$ for \emph{EP}/WXT and $\sim34\%$ for \emph{SVOM}/ECLAIRs.
These values increase to $\sim28\%$ and $\sim68\%$, respectively, at $z>16$, where the
contribution of Pop III stars to the total SFR peaks. These significant fractions
indicate that the detection of Pop III GRBs at high redshift is a realistic prospect.
Extrapolating from these values, we expect that out of 10 GRBs at $z>10$ detected by
\emph{EP}/WXT (\emph{SVOM}/ECLAIRs), roughly one (three) would be a Pop III GRB.

The MSN and RSN models predict Pop II/I GRB detection rates consistent with those
in the VMSN model ($\sim2.4\,\mathrm{events\,yr^{-1}}$ for \emph{EP}/WXT
and $\sim0.9\,\mathrm{events\,yr^{-1}}$ for \emph{SVOM}/ECLAIRs at $z>6$). Furthermore,
these models exhibit a systematic increase in the detectable Pop III GRB fraction---a trend
that aligns with the rising Pop III SFR density across the VMSN, MSN, and RSN model sequence.
The contribution of Pop III GRBs increases with redshift across all IMF models,
making any GRB detected at $z>16$ likely to originate from a Pop III progenitor \citep{2011MNRAS.416.2760C}.
While consistent with this broader picture \citep{2011MNRAS.416.2760C} and building on
the simulations of \cite{2017MNRAS.472.3532M}, our work provides the first instrument-specific
predictions for Pop III GRB detectability with \emph{EP} and \emph{SVOM}. By leveraging
the unique soft X-ray capabilities of \emph{EP}/WXT and \emph{SVOM}/ECLAIRs, we deliver
a concrete observing strategy and clear detectability forecasts. Consequently, our results
serve not only to confirm previous insights but also to provide an essential guide for
forthcoming observational efforts in time-domain astrophysics.

Note that \emph{EP} cannot directly measure redshifts, as it operates only in the X-rays.
Since the requisite optical/near-infrared (NIR) follow-up observations for redshift confirmation
may not be routinely available, the confirmed number of high-$z$ GRBs
is likely to be lower than our estimates. \emph{SVOM}, however, is specifically equipped to
address this challenge. Its dedicated follow-up telescopes enable the prompt identification
of high-$z$ candidates for subsequent NIR spectroscopic observations (see
\citealt{2024A&A...685A.163L}). A non-detection in both bands of the onboard
Visible Telescope, while ambiguous (as it may also indicate a dusty GRB), serves as
a key indicator for high-$z$ candidates. These can then be fast-tracked
for ground-based NIR observations, leveraging accurate positions from \emph{SVOM}'s
Microchannel X-ray Telescope. Looking beyond \emph{EP} and \emph{SVOM}, several missions
dedicated to high-$z$ GRB detection, including \emph{THESEUS}
\citep{2018AdSpR..62..191A,2021ExA....52..183A,2021ExA....52..277G}, \emph{Gamow Explorer}
\citep{2021SPIE11821E..09W}, and \emph{HiZ-GUNDAM} \citep{2024SPIE13093E..20Y}, are being
actively developed. These current and forthcoming missions could therefore offer
a promising and realistic pathway toward the first detection of Pop III stars via
their high-$z$ GRB signatures.

\section*{Acknowledgments}
We would like to thank the anonymous referee for helpful comments.
We are grateful to Tatsuya Matsumoto for providing us with the compiled data set of cosmic SFR density.
This work is supported by the National Key R\&D Program of China (2024YFA1611704),
the Strategic Priority Research Program of the Chinese Academy of Sciences (grant No. XDB0550400),
and the National Natural Science Foundation of China (grant Nos. 12422307, 12373053, and 12321003).



\end{document}